\documentclass[12pt]{article}
\usepackage[english]{babel}
\usepackage{layout}
\usepackage[latin1]{inputenc}
\usepackage{latexsym,amssymb,amsmath}
\usepackage{amsfonts,amsbsy}
\usepackage{eufrak}
\usepackage[mathscr]{eucal}
\usepackage{verbatim}
\usepackage{amscd}
\usepackage{indentfirst}

\begin{document}

\title{\bf {\large{ON THE SCALES OF MASSES IN ELEMENTARY PARTICLES}}}
\author{\normalsize{Luis J. Boya\footnote{\texttt{luisjo@unizar.es}}} \\
\normalsize{ and } \\
\normalsize{ Cristian Rivera\footnote{\texttt{cristian\_elfisico@hotmail.com}}} \\
\normalsize{Departamento de F\'{\i}sica Te\'{o}rica} \\
\normalsize{Universidad de Zaragoza} \\
\normalsize{E-50009 Zaragoza, SPAIN}}
\date{}
\maketitle

\begin{abstract}

We make an attempt to describe the spectrum of masses of elementary
particles, as it comes out empirically in six distinct scales.
We argue for some  rather well defined mass scales, like the electron mass: it seems to us that there
is a minimum mass associated to any electric charge, so we elaborate on     this
assumption; indeed, some scales of masses will cover also masses of
composite particles or mass differences. We extend some plausibility
arguments for other     scales, as binding or self-energy effects of the microscopic
forces, plus some speculative uses, here and there, of gravitation. We also
consider briefly exotics like supersymmetry and extra dimensions in
relation to the mass scale problem, including some mathematical arguments   (e.g. triality),
which might throw light on the three-generation problem. \\

The paper is rather tentative and speculative and does not make
many predictions, but it seems to explain some features of the
particle spectrum.

\end{abstract}

\small{PACS: 06.20 Jr, 06.30 Dr, 12.10 Kt

}

\section{Motivation}

One of the most unsatisfactory features of our understanding of the microworld is the status of the
spectrum of masses: the masses of elementary  particles are not predicted at all, and in the
Standard Model (SM) they are just given by arbitrarily variable couplings to the overall scalar Higgs boson,
undiscovered so far; the coupling is just adjusted as to reproduce the experimental mass:
and this, of course, is none an explanation! \\

For the admittedly large predictive power of the theory of SM one needs first to put by hand the
masses and the coupling constants, as well as some information on the types of acting particles,
like spin, charge, etc. Then many scattering processes, plenty of decay constants and some bound states
can be accurately predicted by the theory: The three known microscopic forces can be described successfully
by the respective gauge theories, and in the three cases many checks can be performed, and are fairly
well borne out by the experiments; it is only when one asks questions about the mass spectrum or the
range of the coupling ``constants'', that the answers are scarce,
or in cases nonexistent at all; indeed, the total number of parameters to be fixed beforehand
to compare experiments with theory is rather large, well beyond twenty \cite{VOI}.
Of course, low-energy calculations in strong interactions (Quantum Chromodynamics, QCD)
are marred for our inability to perform non-perturbative calculations, but even there some successes
(e.g., for many hadrons as bound states) have been achieved by lattice calculations, etc. \\

However, we notice that the particle mass spectrum is not completely chaotic, and some levels and groupings
are clearly phenomenologically apparent. In the present essay we look at the problem of identifying these levels,
and provide, when possible and sensible, of a \emph{rationale} for them.
These groupings might include also masses for some composite particles, e.g. the pion mass or the
neutron-proton mass differences will be considered in some of the mass scales we shall discuss. \\

One of our tenets will be the interpretation of the electron mass scale \linebreak  $\approx$ O(1 MeV),
with a \emph{minimum} mass
supporting particles with electric charge. \\

For a recent alternative use of the Higgs scalar(s) in the SM see \cite{DUE}.\\

The coupling constants are also used as given, but some
speculations based on running towards Grand Unification are also
contemplated, as well at some appeals to extra dimensions and/or
supersymmetry.

\section{The scales of masses: general discussion}

If we look at the experimental masses of particles around us,
they clearly gather in some groups. Here we give just a broad introduction to the subject,
with a specific discussion of each level later on. \\

We observe neatly \emph{six mass scales} (see e.g., Particle Data Group \cite{PDG}):\\

\begin{enumerate}
\item \underline{Massless} particles, $m = 0$. As far as we know, the following particles

\begin{equation}\label{eq:1}
    \textrm{Photon}\, \gamma, \quad  \textrm{Gluon} \;  \textrm{$g$}, \quad \textrm{and} \quad  \textrm{Graviton} (?)\; \textrm{$h$}
\end{equation}

seem to be massless to a large precision (e.g. $m_{\gamma} < 1 \times 10^{-18}$ eV, \cite{PDG}).
In theory, the gluon mass is zero; the graviton is yet to be found, but it is expected to be massless also.

The next level is the neutrino mass scale: although only square differences are measured so far for neutrinos,
there is some consensus on two neutrino mass difference values and the corresponding mixing angles; and there is
no reliable information for a third mass, the third mixing angle being rather small.
The PDG quoted values for the masses are as follows: \\

\item
\underline{Neutrino mass scale}

\begin{equation}\label{eq:2}
   |m_{2}^{2} - m_{1}^{2}|\approx(9 \times 10^{-3})^2  eV^2 \quad    \textrm{and}  \quad
   |m_{3}^{2} - m_{2}^{2}|\approx(4 \times 10^{-2})^2  eV^2
\end{equation}

At a value more than a million times higher it does show up the electron mass scale, around the MeV:

\item
\underline{Electron mass scale}; besides the electron $e$, we include in this level also the first-generation
quarks $u$, $d$:
\begin{equation}\label{eq:3}
\textrm{electron,} \,\, m_e = .512\,\textrm{ MeV};\quad \textrm{quarks:}\, \,  m_u \approx 2  \,\textrm{ MeV},
\, m_d \approx 4\textrm{ MeV}
\end{equation}

Of course, quarks masses (\emph{current} masses for $u$, $d$) are
\emph{deduced}, by a somewhat indirect way, from several
experimental pieces of data; see e.g. \cite{WEIN}, \cite{WEIN A}.

The muon lepton $\mu$ was a fully unexpected surprise when discovered
(1937); today the muon mass \emph{level} is well populated,
with the strange quark $s$, the composite pion $\pi$, the
so-called QCD scale, $\Lambda_{QCD}$, etc.; all these masses
are around 100-200 MeV:

\item
The \underline{muon mass scale}: it includes also the pion, although is not elementary,
and the strange quark $s$:

\begin{equation}\label{eq:4}
 m_{\mu} = 106\,\textrm{ MeV};\quad (m_{\pi} = 137\,\textrm{ MeV}) \quad  m_s \approx 104\textrm{ MeV}
\end{equation}

Around $\Lambda=\Lambda_{QCD}\approx$ 200 MeV, the scale of QCD, the regime changes, roughly speaking,
from asymptotic freedom $(q^2 >> \Lambda)$ to confinement $(q^2 << \Lambda)$.
\item The \underline{nucleon mass scale}: again, proton $p$ and neutron $n$ are not elementary, but the
charm meson $c$ is included, as well as the third lepton, $\tau$, and the bottom quark $b$; all group around the GeV scale:

\begin{equation}\label{eq:5}
\begin{aligned}
&\textrm{$c$ (charmed meson),} \, m_c \approx 1.27\textrm{ GeV}.\, \\
&\textrm{$b$ (bottom meson)}\,
m_b \approx 4.2\textrm{ GeV}.
\end{aligned}
\end{equation}

\begin{equation}\label{eq:6}
\textrm{Tau lepton $\tau$}\,  \textrm{with}\,   m_{\tau}=1.8\textrm{ GeV}.
\end{equation}

\begin{equation}\label{eq:7}
\begin{aligned}
&\textrm{(Proton $p$), as }   m_p= 939\textrm{ MeV}.; \\
&\textrm{(Neutron $n$), as} \, m_n - m_p\approx1.2\textrm{ MeV}.
\end{aligned}
\end{equation}

Finally, we have the electroweak mass scale, with the massive
gauge bosons:
\item The \underline{electroweak (broken) mass
scale}: $W^{\pm}$ and $Z$ vector mesons, as carriers of the
(electro-) weak force, rank at the next level, with masses
around 100 GeV;
also $\langle H \rangle$, the expectation value of the (original neutral,
scalar) Higgs field $H$, is in the same ballpark. The value
of the original (1934) Fermi coupling constant $G_F$ (with
$G_F^{-1/2} \approx$ 292 GeV) was of course also
comparable. The last discovered quark (1995), the top $t$,
is placed also in this level.
Hopefully the new-to-be-discovered Higgs particle(s) would have a mass on the same range, so we have

\begin{equation}\label{eq:8}
 m_{W^{\pm}}= 80 \textrm{ GeV};\quad \quad m_Z= 91 \textrm{ GeV};
\end{equation}

\begin{equation}\label{eq:9}
 m_t= 172\textrm{ GeV};\quad \quad m_H> 114\textrm{ GeV}; \quad <H>= 247\textrm{ GeV}
\end{equation}

With Supersymmetry (Susy) one needs more than one Higgs, but the minimum mass quoted is around the cited limits;
see later.

\end{enumerate}

\underline{Interactions}. These are the clear-cut mass scales we
see experimentally; they group ostensibly in the six
above-mentioned scales. Now the question of \emph{interactions}
arises,
as physically masses should come from forces, from interactions. There should therefore be relations
between \emph{masses} and \emph{forces}. About the forces present in physics,
we take the conventional view of the \underline{four} interactions: Einstein's \emph{general relativity} as a
theory of (pseudo-)Riemannian space-time (with $-+++$ signature), with the geometric description of the
gravitation force: geodesic motion for test particles in a given gravitational field, and curvature generated by
matter as in Einstein equations of gravitation (1915).
Of course, due to the weakness of gravitation on the ordinary microscopic scale,
we can take now as the spacetime manifold just Minkowski space, which is flat.
Nevertheless, gravitation is an essential part of the whole of physics, so one would not be
surprised if it enters somehow also into the microworld, as least as an ordering parameter. \\

And there are three \emph{microscopic forces}, described as gauge
theories, that is,
mathematically as connections in some vector bundles, with the structure group being the composite (non-simple)
Lie group $G = SU(3)_c \times SU(2)_{wi} \times U(1)_Y \equiv (3, 2, 1)$ ($c$ for colour and $wi$,
$Y$ for weak isospin and -hypercharge) and the associated principal and vector bundles. Of course, the group $G$ by itself implies only the
existence of the $8+3+1=12$ gauge vector bosons with ``spin'' or helicity: $s = 1 = |h|$, in the
adjoint representation of the gauge group $G$, and physically massless \emph{if} there is no spontaneous
symmetry breaking (but see again \cite{DUE}), which seems to be the case for colour $SU(3)_c$ and
for electromagnetism, $U(1)_{em}$: the latter is a subgroup of the $SU(2)_{wi}\times U(1)_Y$ group,
the precise ``location''  being measured by the Weinberg angle $\theta_{W}\approx 29^{\circ}$.
The matter contents are the fundamental (or vector) \emph{representations} of the groups: quarks and leptons,
but there are more possibilities; the spin of the matter particles is not predicted,
but it is $s = 1/2$ overwhelmingly; we do not know why. The putative Higgs(es) would have spin zero. \\

It is perhaps interesting to quote here Witten's analysis \cite{WIT}
for the dimension of the natural \emph{internal} spaces acted upon
by the group (3, 2, 1) of the SM: it has to be 7-dimensional, so
here there is an argument for a total of $4+7=11$-dimensional
spacetime (no longer flat), the same dimension to support maximal
supergravity, to be considered briefly later, which also lives in
11 dimensions! \cite{CJS}.  The group-theoretical favourite space is
the homogeneous space $CP^2\times CP^1\times RP^1$, or  $
[SU(3)/U(2)]\times [SU(2)/U(1)] (=S^2 =CP^1) \times [S^1]
(=RP^1)$. \\

However, in the modern M-theory (1995) \cite{WIT 1}, living also in 11 dimensions with extended objects,
an outgrowth of superstring theory, the compactification space might be quite different:
for example, a common choice is a 7-dimensional $G_2$ holonomy manifold \cite{PPT}.
Moreover, as the electroweak group $SU(2)\times U(1)$ is broken down to the $U(1)_{em}$ group,
the homogeneous space is just $S^3$, 3-dimensional as the last part of the former space, namely $S^2 \times S^1$.
Does electroweak breaking have something to do with 11-dimensional space? With maximal supergravity? \\

Summing up, we see the particle spectrum spread out in six levels,
roughly speaking as (1):\, $m_{\gamma}= 0$; (2): $m_{\nu}\approx
10^{-2}$ eV; (3):  $m_e \approx$ 1 MeV; (4): $m_{\mu}\approx$  100
MeV; then (5): $m_{c}\approx 1$ GeV;  and finally (6): $ M_Z
\approx$  100 GeV.
The known four forces seem to be, at first sight at least, at a loss to explain these mass levels;
although level (3) seems dominated by the e.m. forces, and perhaps the (6) scale is due
to (electro-)weak force (breaking)(?).  \\

With this information as input, we want now to see whether some
rational explanation(s) can be advanced for these mass levels, and
for the particles they encompass.

\section{The Massless level}

The \emph{massless} property of the \underline{photon} $\gamma$ is true experimentally to an astonishing
degree, $m_{\gamma} < 10^{-18}$ eV, so Coulomb forces fall off exactly with the $1/r^2$ law; also the photon seems
to be exactly electrically neutral $(q_{\gamma} < 5\times 10^{-30}$ e  \cite{PDG}).
We understand this, as the photon is the carrier of the e.m. force, with  $U(1)$
as gauge group, and the group being abelian, the adjoint representation is the trivial one, so $\gamma$ is
chargeless, and as the $U(1)$ gauge group it is neither spontaneously nor explicitly broken, the $\gamma$
remains massless. \\

The \underline{gluons} $g$ are the carriers of the (colour) strong
force, whose gauge group is $SU(3)_{colour}$, so there are eight =
$3^2-1$ of them; they have not been seen isolated, but only indirectly;
though all studies imply also that the QCD gauge group $SU(3)$ is
exact, so the gluon $g$ must be also massless (but coloured). Now
the continuation of the proven \emph{asymptotic freedom} property
of strong QCD forces (that is, the UV limit $q^{2}\rightarrow\infty$
is trivial, it is a free theory; this ``justifies'' that the
colour self-energy of gluons or quarks generates \emph{no} mass
for them!) will perhaps imply \emph{infrared slavery} \cite{GG}, so
confinement will hide the true masslessness property of the
gluon \cite{WEIN B}. Experimentally, a mass of a few MeV for gluons
cannot be ruled out as today \cite{PDG}. Contrary to photons, which are
chargeless, the gluons carry colour (with the dim-8 adjoint
representation of $SU(3)_c$, as said); so it must be anticipated
that some consequences of the colourful gluons like gluonium
``atoms'', ``glue'' contribution to the mass of hadrons (see
below) etc., will show up eventually. \\

Speculations for the $SU(3)_c$ group as coming from the octonion
numbers are also sometimes
contemplated \cite{RMD}, \cite{BOY 1}: $SU(3)$ is the stabilizer or ``little group'' of the octonion-algebra
automorphism group $G_2$, acting on the $S^6$ sphere of unit imaginary octonions. Also,
manifolds with $G_2$ holonomy, as said, are the favourite ones for compactifying from 11 to our
mundane 4 dimensions \cite{PPT}; in any case, it is just remarkable that the $SU(3)$ group appears at
least \emph{three times} in the phenomenology/theory, to wit: colour, flavour (i.e., the original $SU(3)$
of Gell-Mann and Ne'eman, 1961), as well as the holonomy group of the heterotic string
compactification Calabi-Yau (CY) space. \\

The ``graviton'' $h$ has never been found, and reasonable doubts
exist (e.g. by F. J. Dyson \cite{DYS},  \cite{TON}) it never will; but we
take the conventional view that the long-range decay of
gravitation, i.e. the $1/r^2$ gravitational force law, will
``translate'' into the massless character for the putative
graviton also. The natural mass for any gauge boson is zero,
unless the gauge group is broken; there seems to be no reason why
the $U(1)_{em}$ group should be broken, neither the very same
Lorentz group $L_0$ should be spontaneously broken
(\emph{explicit} breakings of Lorentz invariance are also
contemplated nowadays, but do not take stand in the issue; see
e.g. \cite{COR}). We all hope that at the end of the day the
gravitation interaction should join the other microworld forces,
but at the moment there is a clear-cut distinction; some ideas
along an unification line-of-thought
will be presented as we go along. \\

So the only gauge symmetry broken is the $SU(2)$ group of weak isospin ($wi$); more precisely,
that part of $SU(2)_{wi}\times U(1)_Y$ that leaves the mixed $U(1)$ group of e.m. as an exact symmetry,
the mixing being determined by the Weinberg angle $\theta_W$ \cite{WEI-67}. As consequence, we
have the three massive boson states: $W^{\pm}$ and the neutral $Z$. In our philosophy
that \emph{ any electrically charged particle must have a mass}, we realize why $SU(2)_{wi}$
cannot be exact: the $W$'s are charged. Of course the arguments does not tell about the magnitude of the breaking,
and experimentally the expectation value of the Higgs field, $<H>$, on the 200-GeV range,
is a factor of $10^5$ of the minimal mass scale to support an electric charge:
The reason why also the $Z$ is so massive and why the mass is not the minimum e.m. mass,
like the electron mass, is unclear at the moment: it should come probably from some argument intrinsic to the e.w.
force breaking mechanism: The $W$ must of course be massive, because it is charged, and it is,
but here the charge does not determine the mass, as it seems to be the case for the electron and first-generation
quarks. What about the $Z$ mass? It is clear that the whole weak-isospin triplet $(W^{\pm}, Z)$ is broken
symmetrically, so $m_W \approx m_Z$ is not unexpected. \\

So we believe we throw a little light on the necessity of the breaking of $SU(2)_{wi}$,
and in the exact nature of the both the $U(1)_{em}$ and the $SU(3)_c$  gauge groups $\ldots$

\section{The neutrino mass scale}

The story of neutrinos is worth recalling briefly in our context
\cite{KAB}: first hypothetized as neutral particles and with a tiny (if
at all) mass by Pauli (unpublished) in 1930, they were
instrumental in Fermi's successful ``four-fermion'' beta decay
 theory (1934) \cite{F}; even Fermi already asked himself
about the neutrino mass. When parity violation was discovered in
1957 (supposedly conjectured by Lee and Yang, 1956; decisive
experiments starting by C. S. Wu in January, 1957), the
two-component neutrino theory of H. Weyl (1929) was resurrected to
``justify'' parity violation, in the models of Salam, Landau and
Lee-Yang (1957); neutrinos still entered massless in the
``universal Fermi interaction, $V-A$'', of Sudarshan (1955) and
Feynman---Gell-Mann (1958). To recall that neither Fermi's
original treatment nor the parity-violation refinement of Lee and
Yang dealt with renormalizable theories \ldots  by exactly the same
argument that gravitation was not, namely the appropriate coupling
constant has length dimensions; in fact $[G_N] = [G_F] =
(Length)^2$.
Besides the specific $V-A$ form of the theory, the main advance of this post-war period was the
extension of the original beta decay theory to the whole world of weak interactions,
including muon decay and capture, decay of strange particles, etc. B. Pontecorvo \cite{PON} seems to be
about the first person to conceive unified weak interactions as the natural extension of nuclear beta decay,
around 1947. \\

Two \emph{different} neutrinos $(\nu_e \ne \nu_{\mu})$ were first recognized/identified in 1962,
but the issue of the neutrino masses did not arise experimentally until the turn of the century,
with the ``solar missing neutrino problem'' (see e.g. \cite{BAC}). After some troubles, neutrino(s)
were adjudicated undoubtedly positive mass \emph{differences} around the year 2001 (atmospheric
neutrinos \& Kamiokande experiments, \cite{KMK}); a third neutrino had also been identified nowadays.
In fact, only \emph{squares} of neutrino mass \emph{differences} were measured, with the values quoted at
the beginning, which contain large errors. For an update of the neutrino masses and mixing angles, see \cite{VAL}. \\

Massive neutrinos raise many questions; one is the following: in the late fifties,
Weyl neutrinos were presented as a \emph{rationale} for parity violation, as they were intrinsically
left-handed (hence massless). Then, one might ask, what happens to this argument, that massless neutrinos were
instrumental in ``explaining'' parity violation, once neutrinos have mass?
For a short discussion of this see e.g. \cite{BOY 2}. There is also some speculation about the neutrino mass
differences as a generation effect \cite{REF}, so perhaps the tau-type neutrino would have different mass scale that
the other neutrinos \ldots \\

The standard model SM, conceived since 1970 and completed around 1975, still supposed massless neutrinos \dots  But in
fact, a slight enlargement of the SM will accommodate massive neutrinos without too much trouble. \\

Some actual questions about neutrinos are, for example: \\

\begin{enumerate}

\item
 What determines the small scale, $\approx 10^{-2} -
    10^{-3}$ eV, for some neutrino masses? We have no clue,
    but we offer here the following \emph{negative} argument:
    nature works with the axioms of a \emph{totally compulsory
    (fascist) state}: all which is not forbidden is mandatory;
    there is nothing to impose zero mass for the neutrinos (as
    there is for the photons!), hence neutrinos have to have a
    mass! As they have no charges, the mass could be less than
    the electron mass (and it seems to be!). On the positive
    side, we expect that once gravitation forces will be
    accommodated with quantum mechanics (see later), a kind of
    gravitational and/or weak interaction self-energy of the
    neutrinos (they have weight, after all!) could generate a mass
    for them. That is, as neutrinos experience the (purely)
    weak force, a self-mass is not to be ruled out, with
    origin ``similar'' to electron mass or first quark masses.
    For a clear-cut ``gravitation neutrino'' see \cite{GOL}.

\item  Are \emph{the three} neutrinos massive? Are they more
    than three? At the moment only two mass differences do
    exist, but we believe (and predict, really) that the three
    neutrinos have \emph{no} reason to be massless, hence the
    three of them \emph{must be massive} \cite{MAGB}\ldots  and as the
    reasons should be similar, the three of them should have
    masses in the same range, meV for example, (massive
    $\approx \,  \ge$ 1 eV interacting neutrinos are to be
    excluded by astrophysical reasons); but see \cite{REF}.
    Experimentally, direct measurements of neutrino masses are
    still out of question, but it might come up to be possible
    in the future (for example, after careful measurements of the
    end-spectrum of some nuclear beta decay processes like
    tritium decay, double beta decay (neutrinoless or
    neutrinoful), etc.).

\item  On the other hand, neutrino masses apparently do not
    experience the ``generation effect'' present in other
    leptons and in quarks: electron, muon or tauon have very
    different masses, and so have e.g. the up $u$, the
    charmed $c$ and the top $t$ quarks, as well as $d$, $s$ and $b$. So
    there must be a generation effect, perhaps related to
    charges, which is not (?) present in neutrinos, and which
    we do not understand yet; but again, this is not all
    clear-cut.
\item How do neutrinos mix? The Cabbibo-Kobayashi-Maskawa
    (CKM) matrix for flavour mixing suggests a corresponding
    neutrino mixing matrix, which does exist, but at the
    moment is incompletely known. Although the third mixing
    angle should be rather small, if nonzero, as expected, it
    will allow for an extra $U(1)$ phase contribution  to the $CP$
    violation, rather welcome, to explain the
    matter-antimatter asymmetry present in the actual
    Universe!
\end{enumerate}

For speculations about the masses of the three neutrinos, see \cite{FRITZ}.\\

Are there other hints for the existence of a neutrino scale, turned out in mass,
to be that small? Yes, there are \emph{cosmological} arguments:
(i) The existence of a positive cosmological constant $\Lambda$, producing \emph{accelerated}
cosmic expansion on top of Hubble's constant-velocity flow, is out of question since the year 2000,
and its value translates into the meV scale, close in fact, to the neutrino mass scale \cite{PDG}.
And (ii) besides, the average density of energy in the Universe should also be in this range,
as the cosmological constant amounts for about $.7$ of the mean density of the actual Cosmos \cite{L}.
As the evolution of the Universe is most likely consequence of gravitation, one sees another hint, perhaps,
that the neutrino masses should be related to the gravitation-dominated actual evolution of the Universe
as a whole \cite{SEG}. \\

The neutrinos are still very mysterious. Are they Dirac or
Majorana particles? A particle of type [$m > 0$, $s = 1/2$] has
four components, interpreting negative energies as antiparticle
states; but the neutrinos active in beta decay are fundamentally
chiral (that is, the beta-decay neutrino is left-handed, as if it
were a massless fermion, and the antineutrino would be
right-handed). What about the other two degrees of freedom?  There
is the famous \emph{see-saw} mechanism of Gell-Mann and Ramond
\cite{GMR}, which relates neutrino masses, electroweak breaking scale
and the Grand Unified Theories (GUT) mass scale. Leaving for later
speculations on the GUT scale, the bland argument in \cite{GMR} is that
the electroweak scale (around 100 GeV) is $\propto$ to the
``square root'' of the GUT scale times the actual neutrino scale,
to wit:

\begin{equation}\label{eq:10}
\begin{aligned}
&M_Z^2 /(m_{\nu}\times M_{GUT}) \approx 1, \textrm{ e.g. with}\,  M_Z =90 \textrm{ GeV} \\
&m_{\nu}\approx10^{-2}\textrm{ eV if}\,  M_{GUT}\approx10^{16}\textrm{ GeV}
\end{aligned}
\end{equation}

We do not see how compelling is this see-saw mechanism. \\

Another line of argument, with the same conclusion is perhaps more
cogent: start with the cosmological constant value $\Lambda$
(expressed as an energy); in the future it must be related (at
least) with gravitation; now neutrinos undergo gravitation forces,
so there is no big surprise (?) if both effects are in the same
ballpark, let us say the mili-eV regime \ldots For a recent study
of the Cosmology at the meV-scale, see \cite{MAS}.

\section{The Electron Mass Scale}

We quote first some data \cite{PDG}:

\begin{equation}\label{eq:11}
\begin{aligned}
&\textrm{ electron mass:} \, m_e \approx 0.511 \textrm{ MeV} \\
&\textrm{( with
precision} \pm 13 \textrm{ meV, better than } 10^{-7})
\end{aligned}
\end{equation}\\
\begin{equation}\label{eq:12}
\textrm{ up quark  u,} \, m_u \approx 2.4 \textrm{ MeV;  down quark d, }  m_d \approx 4.8\textrm{ MeV}
\end{equation}\\
The first generation quarks, $u$ and $d$, have large errors in their masses, about 50 \%.
The $d$ is heavier than the $u$ in spite of the charge of the $u$ being twice that of the $d$;
the given masses are understood as \emph{current} masses (as opposed to \emph{constituent} masses,
possible for higher mass quarks). \\

Our philosophy, to repeat, is this: there must be a \emph{minimal}
supporting mass for any electric charge, because the nontrivial UV
behaviour of QED (in modern parlance, QED should be an
inconsistent, ``trivial'' theory); in the conventional,
renormalized theory, the electron \emph{bare} mass is
\emph{infinite}, and everything is computable from the
experimental mass, taken at face value. The empirical electron
mass fixes an electron radius (as expressed already more than 100
years ago by Lorentz (and Poincar\'e)) by the formula $e^{2}/r$
$\approx$ $m_ec^2$  : for $r\approx$ nuclear radius  ( = 2.8
$\times$ $10^{-15}$ m), the mass comes out to be $\approx$ 1/2
MeV. See also \cite{GLG}. \\

Why the $u$ quark (charge +2/3) is lighter than the $d$ (charge -1/3)?
We do not know, we only remark that both masses are bigger than the electron mass\ldots but not much bigger.
Perhaps some subtle unknown QCD argument would explain this mismatch in the future\ldots but in
any case it is satisfactory for us to see that the first-generation quarks $u$, $d$,
have masses just compatible with being electrically charged. \\

Are there other (e.m.?) mass differences of the order of the electron mass? Plenty,
starting by $n$-$p$ mass differences (and also the positive/neutral pion $\pi$ (139.6 MeV
for $\pi^{+}$  vs. 135 for $\pi^{0}$ ). But also, the validity of isospin invariance (Heisenberg, 1932) is
rather good in nuclei, which guarantees that e.g. tritium and $He^3$ have very close masses,
which gives 31 years for the long lifetime of tritium, 18 keV for the reaction energy,
and the best case for limiting $\beta$-decay neutrino mass. We are pretty sure experimentally that,
when a mass difference is assignable to e.m. differences, or to isospin violation in the old language,
then these differences are in the MeV range; this holds equally in elementary hadrons like the $\Sigma$ triplet of
hyperons  as well as in ordinary nuclei. \\

One feature, for example, that comes close to be explained, is that the neutron-proton mass difference
should be positive, as $n$ is, in quark content, ($udd$), $p$ is ($uud$),
and the down quark is more massive than the up one \cite{ACHA}.

\section{The muon scale}

We first recall Rabi's dictum \cite{RAB} ``who ordered that?''
in reference to the very existence of the muon, discovered, as said, in 1937.
For W. Heisenberg, the muon was the biggest mystery of elementary particles \cite{WH}; still today,
the only ``reason'' we see for the existence of (three) generations is from the \emph{anthropic}
point of view: it is $CP$ violation (experimentally unavoidable: this is why we do exist! \cite{KM}),
which require (Kobayashi-Maskawa) at least three quark generations: so the true answer to Rabi's
old question of why muons exist is this \cite{BOY 3}: it was YOU yourself who ordered them, as your very existence
depends on the presence of at least three generations,
muons being part of the second, to explain overabundance of matter vs. antimatter! \\

Unfortunately, it turns out that the measured amount of the $CP$ violation strength
(in neutral $K$ decay, for example) is not enough to explain in quantitative terms
the abundance of matter vs. antimatter in our observable Universe, but it is on the right track.
We expect that the possible $CP$ violation in the \emph{neutrino} mass matrix (see above) should help \ldots \\

As the muon mass is in the same batch as the pion mass (100 MeV vs. 137), one should look perhaps
for a \emph{common mechanism} generating their masses. For the pion $\pi$ there is such a mechanism;
it goes with the name of  ``chiral symmetry breaking'', an emergent phenomenon of strong forces, not
totally understood as today. This global (i.e.  non-gauge) chiral  symmetry (i.e. $SU(2)_L \times SU(2)_R$)
is not shared by the vacuum, and the corresponding Goldstone boson is an hypothetical massless pion,
which becomes massive by some explicit breaking \ldots giving a mass to the $\pi$ much
less than the average hadron masses. We amplify these remarks below. \\

It is remarkable as it is unexplained (but see later for a similar
relation involving the bottom quark $b$ and the $\tau$ lepton) that
the strange quark $s$ and the muon $\mu$ (and also the pion $\pi$)
are in the same ballpark. Also it is remarkable that an ``e.m.''
correction to either pion or muon masses, that is, an
$\alpha$-order correction to the masses (where $\alpha$ =$ e^2/\hbar c$
$\approx 1/137$ is the fine structure constant) gives one back the
electron mass scale! \cite{FRITZ}. For a recent report relating muon mass with many other masses see the essay \cite{MMG}. \\

QCD is a gauge theory of quarks and gluons, with $SU(3)_c$ as the
gauge group. It has been proposed since 1972 (Gell-Mann and
Fritzsch) as the true theory of strong interactions; in this
theory, there is a limit in which one couples massless quarks (the
first generation, $u$ and $d$; it is a worse approximation, but still
viable, with three quarks, adding the strange quark $s$) to the
gluon field; massless helicity $\pm 1/2$ particles can couple the
two helicities differently, as in the weak interaction. Now QCD in
this limit admits however a global (i.e. non-gauge) $SU(2)_L
\times SU(2)_R$  internal invariance group. But this symmetry is
\emph{spontaneously broken} to $SU(2)$ diagonal for some obscure
mechanism  (which we shall not try to select: fermion condensates,
anomaly cancellation, etc., have been proposed as solutions). But
of course, there is then the attendant Goldstone mechanism, as
there are directions from the vacuum which require no effort to
move on: the Nambu-Goldstone (NG) bosons are massless. When this
idea was proposed in the early '60s \cite{GML} it was generally
rejected, because if something was certain in the hadron spectrum
was the absence of massless particles. On the other hand we have
had the pion $\pi$ since 1947, and by mid-1960 it was clearly the
lowest mass hadron, by far: the pion is very light on the hadron
scale, it is pseudoscalar, and carries isospin 1, all consistent
with the way the chiral group is broken, so it may be the NG
boson. Could it be, asked Weinberg \cite{WEIN C} and others, that the
massive but very light pion $\pi$ sould be a reminder of that
spontaneously broken chiral symmetry, which became explicitly
broken by some non-chiral invariant term, giving a little mass to
the pion, which then would become a ``pseudogoldstone'' particle?
One possible explicit chiral breaking term is the quark mass, in
its turn unavoidable in our framework that gives a mass to any charged particle, and the quarks are charged! In
other words, chiral symmetry is broken both spontaneously
as well as explicitly, but we understand the second process (as unavoidable) better! \\

The main question next here is this: is there any theoretical
reason for \emph{that} value for the pion mass? Will it still be
the same (pion mass hundred times the minimal quark mass) for a
QCD without quarks? Is it related to the ``mass gap'' in QCD, one
of the Clay Mathematics Institute problems \cite{CLY}? In all QCD
treatments the chiral breaking mass scale is put by hand; the idea
is that the flavour group $SU(2)_L \times SU(2)_R$ breaks
spontaneously to $SU(2)_I\equiv SU(2)$ diagonal; as said, the
consequential massless boson (Nambu-Goldstone) is the pion;
explicit breaking should account for the $u$, $d$ quark masses, and
also for the very pion mass,
much bigger. Lattice calculations with QCD account for many hadron masses, once the input is given, namely:
the light QCD scale, around 100 MeV, and also the first generation quarks masses, around a few MeV \cite{VAZ}. \\

The same scale is present also in the $s$, the strange quark mass, the third quark to be
discovered (strange particles discovered in the late forties in cosmic rays (Rochester and Butler);
interpreted as the need for the third, strange quark around 1962, with Gell-Mann ``$SU(3)$ flavour symmetry'');
this symmetry is rather badly broken, so it is much poorer than isospin. We know today,
since the old arguments of Glashow \emph{et al}., \cite{GIM} that quarks and leptons should accommodate in the same
generations, lest we confront too much neutral currents with change of flavours.
In particular, the fourth quark, the charmed one $c$, was predicted once the $SU(3)_{flavor}$ group
became accepted, even approximately. Y.Ne'eman was one of the first \cite{NEE} to try to relate the strange quark
$s$ with the $\mu$ lepton, unsuccessfully we must say. There is also an additional anomaly cancellation condition,
first put forward by \cite{BOU}. \\

What is the reason for this intermediate scale? Granted we do not really fully understand any scale,
this level, 100 MeV, is perhaps the most mysterious of all (that is, one can associate e.g. the electron,
proton or $Z$ scale to self-energy or binding effects of the e.m., strong or broken weak force).
So it comes as a partial relief when we notice that QCD exhibits a range of phenomena around
the so-called $\Lambda_{QCD}$, close to 220 MeV. In particular, QCD is a conformal (scale-free) theory,
where the phenomenon of \emph{dimension transmutation} takes place: the dimensionless coupling constant
$\alpha_{QCD}$ is ``traded'' for a renormalization energy scale, that we can identify
with $\Lambda_{QCD}$. What is the relation with the $s$ quark, or the $\mu$ lepton, or for that matter with the
very QCD theory? \\

We insisted on the electron mass coming from QED selfenergy; this
clearly does not apply to the muon: instead, as Barut, Fritzsch
and others have noticed (see e.g. \cite{KOI}), if the muon scale is a
``natural'' one, the electron mass is seen as an electromagnetic
$\alpha$-order correction: it is a very good adjustment to set

\begin{equation}\label{eq:13}
m_{\mu}/m_{\epsilon}= 1+3/(2\alpha)  \approx  207
\end{equation}\\
If this explanation stands, we shall never be able to deduce \emph{masses}, only mass \emph{ratios}.
Some scale, e.g. $\Lambda_{QCD}$ should be taken for granted.

\section{The nucleon mass level}

The bound states of the strong force are to-day called \emph{hadrons},
name due to L. Rosenfeld, \cite{RF}. They come in two classes: mesons, made out of quark-antiquark
pairs $\bar{q}q$, and baryons, made out of three quarks $qqq$ (or  $\bar{q}\bar{q}\bar{q}$); only $SU(3)_c$
singlets are allowed, because the confining character of the nonabelian gauge force at the IR limit:
colourful states do not appear then as free states.
What about the binding energy due to this colour force?
Although we are not much concerned here with reporting masses of bound quarks,
we can add some considerations. There are at least conceptually \emph{three}
different scales of colour binding energy: \\

 \begin{enumerate}

\item In the broken chiral limit, the pion mass sets the
    minimal scale for colour binding, around 100 MeV. In that
    scale one can put, not only the isotriplet of pions, but
    the whole octet $0^-$ as pseudogoldstone bosons of broken
    $SU(3)_L \times SU(3)_R$ flavor, generated by the three
    lightest quarks, $u$, $d$  \emph{and} s. In fact, in the
    eightfold-way (for $SU(3)_{diag}$) the octet seen from
    the I isospin-SU(2)  subgroup splits in pions $\pi$ (I=1,
    three states), Kaons $K$ (I=1/2, four states) and the
    singlet $\eta$ (I=0), all in the $\le$ 1/2 GeV range,
    consistent with: first generation $\bar{q}q$ mesons with
    NG mass reduction, the $\pi$: mass $< 150$ MeV; the four $K$
    mesons, mass $< 500$ MeV (already the $s$ quark, entering the
    $K$ meson bound states, contributes $\approx$ 100 MeV;
    also, the $SU(3)_{flavor}$ is much more badly broken than
    $SU(2)_{isospin}$). Finally, we have the singlet of the
    eta ($\eta$) particle, with mass 548 MeV: comparable to
    the kaon mass, because the strange quark $s$ enters twice.
    Still there is a ninth $p$-scalar meson, $\eta'$, with a
    mass 958: it is not protected neither by the NG mechanism
    nor by being strangeless: the masss turns out to be
    bigger, but sill $< 1$ GeV.

\item Quark-antiquark bound states,  $\bar{q}q$, but outside
    the NG limit; for example, the spin-1 nonet ($\rho,
    \omega, K^{*}, \phi$): all masses beyond 1/2 GeV, and
    less than 1 GeV, except the $\phi$(1020): centrifugal spin
    1, plus strange content plus absence of NG `explains' the
    masses, at least qualitatively. Then there are other meson
    multiplets, as recurrences, higher spins, etc.

\item Baryons as protons and neutrons are made of \emph{three} quarks; the binding energy
turns out to be bigger, and indeed much bigger than the constituents masses, a situation
totally different of the atoms: in the $H$-atom, the binding energy is 13.6 eV,
whereas the rest mass of $e + p$ is bigger than the GeV.
But most of the nucleon mass is ``binding energy'', and,
in spite of some success with lattice calculations, QCD is still
very far away to compare with the successes of the atomic binding energy calculations\ldots \cite{VAZ}.
Wilcek \cite{WCK} is one of these who rightly pointed out that it is \emph{not} true that the ``mass'' of
the Universe comes mainly from the Higgs, the ``God's particle'',
but from the binding energy of the QCD force\ldots as hidden in the nucleons. \\

\end{enumerate}

But the nucleon mass is no doubt very clearly a new scale,
shared also by the charm ($c$) and the bottom ($b$) quarks. Why is
it the nucleon mass propagated to these two quarks? A total
mystery, it seems to us\ldots  But related (may be) to the
same problem of the $s$ quark mass, ``propagated'' from the
chiral symmetry breaking, and perhaps again connected with the
lepton-quark symmetry generation-wise that we mentioned. \\

Notice also the \emph{oblique symmetry} in the second and third generations: $\{e, \nu_{e}\}$
go with $\{u, d\}$
as the first generation. Then  $\{\mu, \nu_{\mu}\}$ go with $\{c, s\}$ as the second: only the strange quark $s$
appears with $\Lambda_{QCD}$-type mass. And then $\{\tau, \nu_{\tau}\}$  goes with $\{t, b\}$
for the third generation, but only $\tau$ lepton matches with $b$ quark. On top of all this,
the $c$ quarks lies in the same ballpark as the $\tau$ lepton and the $b$ quark, whereas the top quark $t$
goes to the next mass level, the $W^{\pm}$-$Z$ level. Indeed the ``relation'' between $s$ quark and $\mu$
lepton repeats itself with the bottom quark $b$ and the $\tau$ lepton, as a renormalization group effect. \\

\emph{Nuclear} binding energies, opposed to\emph{ quark} binding energies, are small,
if one considers nucleons as composed of three quarks; for
example, the deuteron ($d = p-n$) binding energy is 2.2
MeV, out of 2 GeV rest mass. This is simple, if understood as
a small, ``molecular'' effect. Molecules, in fact, have a
binding energy much smaller than the $H$-atom binding, say
centi-eV against eV. For nuclei, that should be justified
soon, from QCD we hope, and it is expected that lattice
calculations of complex nuclei should account for that
nuclear binding energy \cite{VAZ}. For physicists of the old
generation it came as a surprise when it turned out that most
of the ``nuclear binding energy'' is a sort of molecular or van der Waals residual force \ldots \\

So strong forces, as described by QCD, result in two mass
scales, say $\Lambda_{QCD}$ $\approx$ 150 MeV and $m_N$
$\approx$ 1 GeV, represented e.g. by the pions $\pi$ and the
nucleons $N$; and these two scales propagate to bare quarks and
leptons, as we pointed out. As stressed above it seems that
only $\Lambda_{QCD}$  is primitive, and we should be eventually
able to compute nucleon mass \emph{ratios} from, say, lattice
QCD calculations.

\section{The broken electroweak scale }

 The 100 GeV scale, our next level, is rather well populated: we have here the $W^{\pm}$, the $ Z$,
 the top quark $t$, as elementary particles, and also the Higgs vacuum expectation $\langle H \rangle$,
 plus hopefully the Higgs particle itself, and of course the (old) Fermi coupling constant,
 $G_F$ $\approx$ 298 GeV, traded today by this expectation value $\langle H \rangle$.
 One then asks, what is the geometry of the spontaneous electroweak breaking? How is the vacuum manifold? \\

The same problem as before arises also: granted that for some reason the e.w. break scale mass is
in the 100 GeV range, why it does attach to the top mass (and to the Higgs mass)?
We have very bluntly seen the \emph{generation problem}: each of the three generations defines a mass
scale for quarks (1 MeV the first; 100 MeV ($s$) - 1 $GeV$ ($c$), second; and 1 GeV ($b$) - 100 GeV ($t$) the third),
with quarks lying on that range: $u$ $\&$ $d$ for the smallest, $s$ for the second, $c$ and $b$ for the third, $t$ for the top.
Perhaps the most expected result is a simple relation between the Higgs mass and the Higgs vacuum
expectation value, but even this cannot be checked until the Higgs is discovered.
Summing up, we have a generation effect, as well as an \emph{oblique effect}, and the Higgs participates,
as perhaps a kind of fourth generation\ldots \\

We have again no clue as the e.w. scale; the bare dimensionless e.w. coupling constant is of
the order of the e.m.'s $\alpha$, but the weak interactions are ``weak'' because they are broken,
and the breaking scale is much higher than both atomic and nuclear masses. \\

Among the speculations for the e.w. scale one can contemplate for example Supersymmetry,
Grand Unification or compactification from Higher Dimensions\ldots We shall
say something more on this problem later in this review.

\section{Two more (theoretical) scales: GUT and Susy }

There are no particles found, supposedly elementary, with masses much beyond 100 GeV, although there
are candidates; e.g. Susy partners, very massive see-saw neutrinos, etc.
Empirically we have also the nasty problem of the \emph{dark matter}, constituting about $25\%$ of the mass of the Universe.\\

However, the three \emph{running coupling constants}, respectively for QCD,
$\alpha_{\textrm{QCD}}$ and for e.w. forces $\alpha_{\textrm{EM}}$ and $\alpha_{\textrm{W}}$, by renormalization group calculations,
starting with Georgi, Quinn and Weinberg (1974) \cite{GQW} seem to (roughly) coincide at an enormous scale, $10^{15}$ GeV.
This important calculation points out at least to two items: (1),
Grand Unification Theories, GUT: if the three interactions are equivalent at the energy scale of $10^{15}$ GeV,
one should understand the different values we observe ``at rest'' for the coupling constants as consequence
of the different speed of running of the three coupling constants, which is well understood
from renormalization group arguments.(2) By the way, the matching of the three couplings is much improved
with Supersymmetry, which also extends about an order of magnitude the coincident energy ($10^{15}$ to $10^{16}$ GeV;
as comparison, Planck's mass scale is around $10^{19}$ GeV); the couplings seem to unify at  the value
$\alpha_{\textrm{GUT}} \approx 1/25$.
For a modern treatment of gravitational corrections to the running coupling constants, see \cite{WIL}.\\

The first GUT group historically was $SU(5)$, by Georgi and Glashow, \cite{GG}.
The unifying group has to have complex representations (to account for parity violation,
so fermions and antifermions fill up complex \emph{conjugate} representations (=\emph{ irreps}),
and there are not so many possible groups: only $SU(n)$, $n \ge 3$, $SO(4n+2$) with the spin irreps,
and $E_6$ are the candidates among simple Lie groups; curiously, for a Lie group to have complex
representations one needs the centre of the group to have more than involutive elements \cite{BOY 4},
and indeed a natural hierarchy of GUT groups is $SU(5)$ inside $Spin(10)$ inside $E_6$ with centres $Z_5$, $Z_4$, $Z_3$.
But the matter is not yet mature \ldots It is a bit surprising and uneasy for us to learn that electric
and weak forces were successfully unified back in 1967 (Weinberg), but
in the remaining 40+ years we have been unable to progress any further.
Hints of GUT unification are still lacking, like the much-awaited for proton decay.\\

There is also the famous (already quoted) \emph{see-saw} mechanism of
Ramond \emph{et al}. \cite{GMR}: the neutrino mass times the GUT scale is
about the square of the $Z$ mass \ldots At least they are related. So we
have two or three mostly theoretical arguments for the existence
of a $7^{th}$ scale, around $10^{15-16}$ GeV. The appeal to gravitation is
unavoidable, as the Planck mass scale is not far up (see just
below), but at least this has a merit: by the mentioned see-saw
mechanism, the (very small) neutrino mass scale matches with the
cosmological constant $\Lambda$ (in corresponding units), and it
relates also to the (very large) GUT mass; this makes
``smell'' again of gravitational connotations, not yet
understood.\\

\emph{Supersymmetry} (\emph{Susy}) enters the game now: with the
MSSM, i.e., the minimal Susy extension of the SM model, the
matching of the three coupling constants improve, as said, but to
a larger scale: $10^{16}$ GeV, ten times higher. It is one of the main
reasons why people desire Susy; other reasons are: (ii) the
hierarchy problem: the Higgs should acquire enormous
renormalization mass, unless it has a fermion partner; the Higgs
mass is expected to be less of 200 GeV in any reasonable theory;
(iii) Susy partners are candidates for dark matter, e.g. the
``neutralino'': the dark matter problem arises in
astrophysics, as e.g. the rotation curves of galaxies require much
more mass than the one we ``see''; the dark matter problem,
together with the dark energy issue (which is about the repulsive
acceleration of the Universe expansion) are perhaps the two more
pressing problems to-day in Cosmology and Astrophysics.\\

But Supersymmetry raises more questions that it solves: Susy, if
exists at all, must be broken, and this makes a new scale to
enter: \emph{the scale of Susy breaking}. Below (Sect.11) we elaborate
more on Supersymmetry; at any rate, it might well signal the start
of an eighth scale, perhaps on the TeV range!

\section{The Planck scale}

Gravitation as a whole, as an interaction on its own, has been mainly left out intentionally,
but now it is time to get it back. With $\hbar$, $c$ and $G_N$, we concoct units for everything, in particular
the energy unit is $M_{Pl}$, that is (with $\hbar$, $c$, factors as units) $10^{+19}$ GeV, not too far from the Susy-GUT scale.
What does this mean? We wish we know!\\

We should understand also why the GUT scale is NOT much different from the Planck scale.
Does this lead to a relation between gravity and the other forces? We believe so, in a mysterious way.
One point however we want to emphasize:\\

There is no doubt that the na\"{i}ve yuxtaposition of Quantum Mechanics and General Relativity is wrong:
gravitational interactions are unavoidably not renormalizable, as [$G_{N}] \propto L^{2}$.
As both theories have a clear domain of application, some modification is to be expected, soon or later.
We bet our horses on noncommutative geometry (A. Connes \cite{CNN}), but it is only one of the several proposals
(loop quantum gravitation is another: Ashtekar, 1986 \cite{ASK}; not to speak of superstring theories \cite{GSW} \ldots).
This has been the main reason why we did not consider gravitation as a theory on its own in this review,
except for marginal comments, fixing perhaps an scale, and also influencing, may be, another two.\\

We appeal also to a recent paper by us \cite{BYRS} for the idea of changing the fundamental constants
(in name, not in values). But the Planck mass stands as originally.

\section{A new view on Supersymmetry}

The way to deal with fermions and bosons together in the same pot was accidentally discovered
in the West around 1972, simultaneously with some Russian work, when P. Ramond put fermions in an
incipient string theory. Since 1974 (Wess and Zumino in \cite{FER}) it has been rightly considered as a natural
extension of quantum field theory. Unavoidably, it was considered as a mechanism to
understand features of the real world, in absence of any clear experimental corroboration;
for example, as mentioned, the mass of the Higgs gets unrenormalized to much higher scales if
it has a Susy partner (higgsino); also, the Susy running of the couplings imply different
Higgses for the upper vs. the lower quarks, and it goes some way to understand the mass differences
between lower quarks and leptons for the second and third generation ($s$ vs. $\mu$, and $b$ vs. $\tau$).
There are other blessings as well, that we omit. For the prospects of finding Susy partners with the
LHC machine, see \cite{ROY}. \\

When Susy appeared, it was hard to swallow for the average physicist.
We were used to consider fermions on the fundamental or vector representation of the gauge groups,
whereas gauge vector (spin 1) bosons (\emph{gaugeons}, one might be tempted to say) went with the
adjoint representation; there is no more fundamental physical difference between particles and fields
that the electron, as a fermion, obeying the exclusion principle, that accounts for all the chemistry,
and the photons, with their cooperative states, and the ``likeness'' of photons to stay together
(coherent states in the laser, etc.). But today we perhaps start to understand better the matter,
and the contradiction is not so poignant. Here is a very bold mathematical idea: \\

In precisely eight \emph{space} dimensions (and \emph{only} in that!) spinor and
vector representations are isomorphic: the centre of the $Spin(8)$
group is $V \equiv  Z_2 \times Z_2$, and also the three
representations: vector $\square$ or $8_V$, and the two
spinor \emph{irreps} $\Delta_L$ and $\Delta_R$, are
\emph{equivalent} (isomorphic), as the symmetry group of that
centre, $S_3 = Aut (V)$, lifts to a \emph{true} symmetry of the
$Spin(8)$ group: this is called Cartan's \emph{triality} in
mathematics, and it is very closely linked to the octonion
division algebra; triality is very obvious from the Dynkin diagram
for the $D_4\approx O(8)$ group. On the other hand, the
spin-statistics theorem is \emph{not} valid in 8 dimensions \cite{BYS};
so one can contemplate a spinor(s)-vector \emph{bona fide}
symmetry (not supersymmetry!) which descends to four dimensions,
becoming the usual fermion-boson supersymmetry! The speculation
that this is the origin of supersymmetry down to our mundane,
three spatial dimensions, is a strong one, and we tentatively
subscribe to it. On (possibly) compactification, spinors become
fermions, as we see them, with the attendant exclusion principle.
Of course, the adjoint representation kills the centre (it is a
faithful \emph{irrep} of the $PO(8)$ group, = $Spin(8)/V$), so if
gauge groups appear in the process of compactification (as e.g.
removal of singularities: Achyara-Witten mechanism, \cite{AW}),
conventional boson-fermi partners should appear.

\section{Conclusions}

The particles we believe nowadays considered elementary that one observes in nature
group naturally in six well-defined scales, at least.
The massless scale (1), the electron scale (3) and the nucleon scale (5) as present in two quarks ($c$, $b$)
and a lepton ($\tau$), are sort of understood: exact gauge carriers, support of minimal electric charge,
regular binding energy from strong forces. One can perhaps anticipate some understanding of the necessity
of $\Lambda_{QCD}$, as ``dimension transmutation'' of the scale-invariant QCD coupling by a mass (scale (4)).
The electroweak gauge group has to be broken, as carriers are charged, and this points towards the Higgs scale (6).
Only the neutrino scale (2) is not mentioned, and for it we also advanced some gravitation/cosmological arguments.
But of course, all this is much more a research program that a well established set of (unconnected?) hypotheses.
In particular, we want to finish just to emphasize that the main problems remain as intractable as always:
why are there three generations, with partial but also oblique symmetry?; neutrino masses seem to be insensible
to generations (?), with the lower quarks ($s$, $b$) seemingly related to the charged leptons ($\mu$, $\tau$),
at least in the second and third generation, whereas the upper quarks signal the new scale: the charm quark $c$
points towards the QCD binding energies, whereas the top quark ($t$) mass is in the regime of the e.w. breaking scale.
Dark matter raises its nasty head pointing to another scale, with probably cosmological significance. \\

Some of the facts we have signalled have to be the way they are for
\emph{anthropic reasons}; we already alluded to three generations (at least)
to support $CP$ violation, and the enormous abundance of matter vs.antimatter;
but there are other examples: neutrons heavier than protons are essential to
form hydrogen, and after this, the remaining atoms and molecules.
Related with this is the necessity of spin 1/2 fermions, to make structures via the exclusion principle.
For a recent review of particle masses, with emphasis on neutrinos, see \cite{BET}. \\

To end up, we would like to stress that the actual electroweak gauge
symmetry \emph{breaking} mechanism is rather ugly and \emph{ad hoc}.
At any rate, as we state at the very beginning, the masses obtained in the conventional
SM by couplings to the Higgs are also very unsatisfactory as a matter of principle.\newpage

\underline{Acknowledgements.} This work has been supported by CICYT (grant FPA-2006-02315) and DGIID-DGA
(grant 2007-E24/2), Spain. We acknowledge discussions with several colleagues in Zaragoza.
The first idea of the paper came up in talks with Alex Rivero,
to whom we owe many comments and some references.
We have had further fruitful discussions with our colleagues
here: J.M. Gracia-Bond\'ia, A. Asorey, A. Segu\'i, J.L. Cort\'es and V. Azcoiti. We thank all of them.


\begin{thebibliography}{99}
\bibitem{L}
The critical density of the Universe is \cite{PDG} $\approx$  $1\times 10^{-5} h^{2} (GeV/c^{2}) cm^{-3}$, and $70\%$
is contributed by the repulsive Cosmological Constant.
\bibitem{MAGB}
Some people claim that, as the three neutrinos oscillate, the three must be massive.
\end{thebibliography}
\end{document}